\documentclass[conference]{IEEEtran}
\IEEEoverridecommandlockouts
\usepackage{cite}
\usepackage{amsmath,amssymb,amsfonts}
\usepackage{algorithmic}
\usepackage{graphicx}
\usepackage{textcomp}
\usepackage{xcolor}
\def\BibTeX{{\rm B\kern-.05em{\sc i\kern-.025em b}\kern-.08em
    T\kern-.1667em\lower.7ex\hbox{E}\kern-.125emX}}

\newcommand{\bpf}{\begin{IEEEproof}}
\newcommand{\epf}{\end{IEEEproof}}
\def\mymedskip{\vskip\medskipamount}
\def\mymedbreak{\par \ifdim\lastskip<\medskipamount
  \removelastskip \penalty-100 \mymedskip \fi}
\def\myaftermedspace{\par \ifdim\lastskip<\medskipamount
  \removelastskip \penalty55\mymedskip\fi}

\newenvironment{example}%
{\mymedbreak\refstepcounter{Exc}
                      {\em Example \theExc:}\enspace}%
{\eop\myaftermedspace}
\newtheorem{teor}{Theorem}[section]
\newcounter{Exc}

\newtheorem{defi}[teor]{Definition}
\newtheorem{lem}[teor]{Lemma}
\newtheorem{cor}[teor]{Corollary}
\newtheorem{con}[teor]{Conjecture}
\newtheorem{prop}[teor]{Proposition}
\newtheorem{rem}[teor]{Remark}
\newcommand{\beq}{\begin{equation}}
\newcommand{\eeq}{\end{equation}}
\newcommand{\beql}[1]{\begin{equation} \label{#1}}
\newcommand{\eeql}{\end{equation}}
\newcommand{\beqa}{\begin{eqnarray*}}
\newcommand{\eeqa}{\end{eqnarray*}}
\newcommand{\beqal}[1]{\begin{eqnarray} \label{#1}}
\newcommand{\eeqal}{\end{eqnarray}}
\newcommand{\beqan}{\begin{eqnarray}}
\newcommand{\eeqan}{\end{eqnarray}}

\newcommand{\bex}[1]{\begin{example} \label{#1}}
\newcommand{\eex}{\end{example}}

\newcommand{\cL}{{\cal L}}

\newcommand{\cR}{{\cal R}}
\newcommand{\cS}{{\cal S}}

\newcommand{\cU}{{\cal U}}

\newcommand{\bF}{{\bf F}}

\newcommand{\bfs}{{\bf s}}

\newcommand{\per}{{\rm per}}

\newcommand{\PG}{{\rm PG}}
\newcommand{\PGL}{{\rm PGL}}
\newcommand{\GL}{{\rm GL}}

\newcommand{\PSL}{{\rm PSL}}

\newcommand{\Tr}{{\rm Tr}}

\newcommand{\gs}{\sigma}

\newcommand{\gth}{\theta}
\newcommand{\gw}{\omega}

\newcommand{\tg}{\tilde{g}}

\newcommand{\tgw}{\tilde{\omega}}
\newcommand{\txi}{\tilde{\xi}}

\newcommand{\tT}{\tilde{T}}
\newcommand{\tL}{\tilde{L}}
\newcommand{\tC}{\tilde{C}}

\newcommand{\tnu}{\tilde{\nu}}

\newcommand{\ga}{\alpha}

\newcommand{\gd}{\delta}

\newcommand{\gl}{\lambda}

\newcommand{\tc}{{\tilde{c}}}
\newcommand{\tbc}{{\tilde{\bfc}}}

\newcommand{\cG}{{\cal G}}
\newcommand{\cH}{{\cal H}}

\newcommand{\Gxi}{\mbox{$\langle\xi\rangle$}}

\newcommand{\Gth}{\mbox{$\langle\gth\rangle$}}
\newcommand{\ord}{{\rm ord}}

\newcommand{\bbF}{\mathbb{F}}

\newcommand{\bbZ}{\mathbb{Z}}

\newcommand{\clF}{\overline{\bbF}}
\newtheorem{fact}[teor]{Fact}
\newtheorem{problem}{Problem}
\newtheorem{exercise}{Exercise}

\newtheorem{alg}[teor]{Algorithm}
\newcommand{\ben}{\begin{enumerate}}
\newcommand{\een}{\end{enumerate}}
\newcommand{\bit}{\begin{itemize}}
\newcommand{\eit}{\end{itemize}}
\newcommand{\Tm}[1]{Theorem~\protect\ref{#1}}
\newcommand{\Le}[1]{Lemma~\protect\ref{#1}}

\newcommand{\Sec}[1]{Section~\protect\ref{#1}}

\newcommand{\btm}[1]{\begin{teor} \label{#1}}
\newcommand{\etm}{\end{teor}}
\newcommand{\btmn}[2]{\begin{teor}[#1] \label{#2}}
\newcommand{\etmn}{\end{teor}}
\newcommand{\ble}[1]{\begin{lem} \label{#1}}
\newcommand{\ele}{\end{lem}}
\newcommand{\bpn}[1]{\begin{prop} \label{#1}}
\newcommand{\epn}{\end{prop}}
\newcommand{\bde}[1]{\begin{defi} \label{#1}}
\newcommand{\ede}{\end{defi}}
\newcommand{\bco}[1]{\begin{cor} \label{#1}}
\newcommand{\eco}{\end{cor}}
\newcommand{\bcorn}[2]{\begin{cor}[#1] \label{#1}}
\newcommand{\ecorn}{\end{cor}}
\newcommand{\bcon}[1]{\begin{con} \label{#1}}
\newcommand{\econ}{\end{con}}
\newcommand{\bbFact}[1]{\begin{fact} \label{#1}}
\newcommand{\efact}{\end{fact}}
\newcommand{\bpr}[1]{\begin{problem} \label{#1}}
\newcommand{\epr}{\end{problem}}
\newcommand{\bprnn}[1]{\begin{problemnn} \label{#1}}
\newcommand{\eprnn}{\end{problemnn}}
\newcommand{\bprn}[2]{\begin{problem}[#1] \label{#2}}
\newcommand{\eprn}{\end{problem}}
\newcommand{\bexer}[1]{\begin{exercise} \label{#1}}
\newcommand{\eexer}{\end{exercise}}
\newcommand{\bre}[1]{\begin{rem} \label{#1}\rm} 
\newcommand{\ere}{\end{rem}}
\newcommand{\balg}[1]{\begin{alg} \label{#1}}
\newcommand{\ealg}{\end{alg}}
\newcommand{\bqu}{\begin{question}}
\newcommand{\equ}{\end{question}}
\newcommand{\bs}{\begin{solution}}
\newcommand{\es}{\end{solution}}
\newcommand{\bh}{\begin{hint}}
\newcommand{\eh}{\end{hint}}
\newcommand{\bms}[1]{\begin{multisolution}{#1}}
\newcommand{\ems}{\end{multisolution}}
\newcommand{\bquo}{\begin{quote}}
\newcommand{\equo}{\end{quote}}
\newcommand{\la}{\langle}
\newcommand{\ra}{\rangle}
\newcommand{\PAut}{{\rm PAut}}

\newcommand{\diag}{{\rm diag}}
\newcommand{\comment}[1]{}

\newcommand{\bfaa}{{\bf a}}

\newcommand{\bfc}{{\bf c}}

\begin{document}

\title{Non-standard linear recurring sequence subgroups and automorphisms of irreducible cyclic codes\\
\thanks{This work was supported by the Estonian Research Council grant PRG49.}
}

\author{\IEEEauthorblockN{%
Henk D. L.\ Hollmann}
\IEEEauthorblockA{\textit{Institute of Computer Science} \\
\textit{University of Tartu}\\
51009 Tartu, Estonia \\
henk.d.l.hollmann@ut.ee}
}

\maketitle

\begin{abstract}
%

Let \(\cU\) be the multiplicative group of order~\(n\) in the splitting field \(\bbF_{q^m}\) of \(x^n-1\) over the finite field \(\bbF_q\). Any map of the form \(x\rightarrow cx^t\) with \(c\in \cU\) and \(t=q^i\),  \(0\leq i<m\), is \(\bbF_q\)-linear on~\(\bbF_{q^m}\) and fixes  \(\cU\) set-wise; maps of this type will be called {\em standard\/}. Occasionally there are other, {\em non-standard\/} \(\bbF_q\)-linear maps on~\(\bbF_{q^m}\) fixing \(\cU\) set-wise, and in that case we say that the pair \((n, q)\) is {\em non-standard\/}. We show that an irreducible cyclic code of length~\(n\) over \(\bbF_q\)  has ``extra'' permutation automorphisms (others than the {\em standard\/} permutations generated by the cyclic shift and the Frobenius mapping that every such code has)  precisely when the pair \((n, q)\) is non-standard; we refer to such irreducible cyclic codes as {\em non-standard\/} or {\em NSIC-codes\/}. In addition, we relate these concepts to that of a non-standard linear recurring sequence subgroup as investigated in a sequence of papers by Brison and Nogueira. We present several families of NSIC-codes, and two constructions called ``lifting'' and ``extension'' to create new NSIC-codes from existing ones. We show that all NSIC-codes of dimension two can be obtained in this way, thus completing the classification for this case started by Brison and Nogueira.

\end{abstract}


\section{\label{LSint}Introduction}
The general problem of determining the automorphism group of any cyclic code is a difficult problem, and \cite[Section 3.5]{charhand} suggests to investigate special cases such as irreducible cyclic codes. Usually, the only permutation automorphisms of an irreducible cyclic code are those generated by the cyclic shift and the Frobenius mapping; we refer to these codes as {\em standard\/} and to the exceptional ones that possess ``extra'' permutation automorphisms as {\em non-standard\/}. For briefness, we will refer to non-standard  irreducible cyclic codes as {\em NSIC-codes\/}. Examples include the $q$-ary simplex codes, certain even-weight codes, and the duals of the binary and ternary Golay codes. 
The ultimate goal is to obtain a full classification of the NSIC-codes. We present several families of NSIC-codes, together with construction techniques called {\em lifting\/} and {\em extension\/} to construct new NSIC-codes from existing ones. As one of our main results, we show that 
every NSIC-code of dimension two can be obtained in this way. 

Interestingly, the notion of an NSIC-code can be related to various other notions. In a series of papers, see, e.g., \cite{bn1,bn2,bn3, bn4}, Brison and Nogueira investigated the concept of a {\em non-standard linear recurring sequence subgroup\/} or {\em NSLRS-group\/}, 
a multiplicative subgroup in an extension of a finite field~$\bbF_q$ with the property that the elements can be represented by a non-cyclic linear recurring sequence with characteristic polynomial over~$\bbF_q$ (for precise definitions of these and other notions, we refer to the next two sections). We show that the notion of an NSIC-codes coincides with that of a NSLRS-group in the case where the characteristic polynomial of the linear recurrence relation 
is required to be {\em irreducible\/} over~$\bbF_q$. 

%
Let $\cU$ be the multilicative subgroup of order~$n$ in an extension $\bbF_{q^m}$ of~$\bbF_q$, with $m$ minimal. Usually, the collection $\cL(n,q)$ of $\bbF_q$-linear maps on $\bbF_{q^m}$ that fix $\cU$ set-wise consists of the maps $x\rightarrow ux^{q^i}$ ($u\in \cU, 0\leq i<m$) only. But occasonally, $\cL(n,q)$ contains other, {\em non-standard\/} maps; in that case, we refer to the pair $(n,q)$ as {\em non-standard\/}. We show that an irreducible cyclic code $C$ of length~$n$ over~$\bbF_q$ is NSIC-code precisely when $(n,q)$ is non-standard; in fact, we show that $\cL(n,q)$ is a group and $\cL(n,q)\cong \PAut(C)$, the permutation automorphism group of~$C$.

The contents of this paper are the following. In \Sec{LSnot} we introduce the notation used in this paper. In \Sec{LSlrs}, we sketch the required background on linear recurring sequences and provide definitions of some of the notions used above.
In \Sec{LSauto} we prove the equivalence of NSIC-codes and non-standard pairs. Lifting and extension is discussed in~\Sec{LSlex}. 
In \Sec{LSes} we introduce equally-spaced polynomials and a related class of NSIC-codes based on a new type of cyclic product codes. \Sec{LScases} lists the NSIC-codes known to us, and in~\Sec{LSclass} we classify the NSIC-codes of dimension~2. Finally, in \Sec{LSdis} we discuss our results.

We have space to include just a few complete proofs, for all other proofs, further background, and any unexplained notation,  we refer to~\cite{cycl-subm1} and \cite{cycl-subm2}. Together, these two papers form an improved and extended version of the arxiv paper~\cite{cyclic-subm}.

%
\section{\label{LSnot}Notation and preliminaries}
%
Throughout this paper, $q$ denotes a power of a prime, $n$ is a positive integer such that $\gcd(n,q)=1$, and~$m=\ord_n(q)$, the smallest positive integer for which $n\mid q^m-1$. The finite field of size~$q$ is denoted by~$\bbF_q$, and we use $\bbF_q^*$ to denote both the non-zero elements of the field and the multiplicative subgroup of~$\bbF_q$. We let $\clF_q$ denote the algebraic closure of~$\bbF_q$.
For a subset~$\cH$ of a group~$\cG$, we write $\cH\leq \cG$ to denote that $\cH$ is a subgroup of~$\cG$. 
There is a {\em unique\/} multiplicative subgroup in~$\clF_q$ of order~$n$, consisting of the $n$-th roots of unity in~$\clF_q$, which we denote by~$\cU_{n,q}$.
Note that $\cU_{n,q}\leq  \bbF_{q^m}^*$, and by the definition of~$m$, $\bbF_{q^m}$ is the smallest extension of~$\bbF_q$ that contains~$\cU_{n,q}$. We let $\xi$ denote a fixed element in~$\bbF_{q^m}^*$ of order~$n$, and we mostly use~$g$ to denote  the minimal polynomial of~$\xi$ over~$\bbF_q$; note that the degree $\deg(g)$ of~$g$ equals~$m$. Then $\cU_{n,q}=\Gxi$, the multiplicative group generated by~$\xi$.

An important notion in this paper is the {\em $q$-order\/}. The $q$-order of~$\xi$, denoted by $\gd_q(\xi)$, is the smallest positive integer~$d$ for which $\xi^d\in\bbF_q$. It is not difficult to show that $\gd_q(\xi)=\gd_q(n):= n/\gcd(n,q-1)$, so depends only on~$n$. 
Writing $d=\gd_q(n)$ and $e=(n,q-1)$, we have $n=de$ with $e\mid q-1$ and $\gcd(d,(q-1)/e)=1$;  as a consequence, the number $m=\ord_n(q)$ depends only on~$d$ and is the smallest positive integer such that $d\mid (q^m-1)/(q-1)$. 
For proofs and more information on the $q$-order, see~\cite{cycl-subm1}.

 
In this paper, we investigate the group~$\cL(n,q)$ of all $\bbF_q$-linear maps~$L$ on~$\bbF_{q^m}$ that fix the group $\cU_{n,q}$ of order~$n$ in~$\bbF_{q^m}^*$ {\em set-wise\/}, that is, such that $L(\cU_{n,q})=\cU_{n,q}$.  Note that $\cL(n,q)$ is indeed a group: since $m=\ord_n(q)$, the $\bbF_q$-span of the elements $1, \xi, \ldots, \xi^{m-1}$ of~$\cU_{n,q}=\Gxi$ is~$\bbF_{q^m}$, so if $L\in\cL(n,q)$, then ${\rm Im}(L)=\bbF_{q^m}$, hence $L$ is invertible. We refer to an $\bbF_q$-linear map on~$\bbF_{q^m}$ of the form $L: x\rightarrow cx^{q^j}$, for some $j\in[m]:=\{0,1,\ldots, m-1\}$ and $c\in\bbF_{q^m}$, as {\em standard\/}. Note that such a standard map is contained in~$\cL(n,q)$ if and only if $c\in\cU_{n,q}$ . We write $\cL_{\rm st}(n,q)$ to denote the subgroup of~$\cL(n,q)$ consisting of the standard maps in~$\cL(n,q)$. In certain exceptional cases, 
$\cL(n,q)$ is stricktly larger than~$\cL_{\rm st}(n,q)$; in that case, we refer to the pair $(n,q)$ as {\em non-standard\/} and to the group $\cU_{n,q}$ as {\em non-standard (over~$\bbF_q$\/}); in the usual case where $\cL(n,q)=\cL_{\rm st}(n,q)$, we refer to both $(n,q)$ and $\cU_{n,q}$ as {\em standard\/}. 

Let $\cS_n$ denote the symmetric group on~$n$ symbols, 
the group of all permutations on~$[n]$. Define the map
$\Psi: \cL(n,q)\rightarrow \cS_n$
by $\Psi(L)=\pi$ if $L(\xi^i)=\xi^{\pi(i)}$ for $i=0, \ldots, n-1$.
It is easily seen that $\Psi$ is a one-to-one group-homomorphism. We will write $\cS(n,q)=\Psi(\cL(n,q))$ and $\cS_{\rm st}(n,q)=\Psi(\cL_{\rm st}(n,q))$ to denote the images in~$\cS_n$  under~$\Psi$ of~$\cL(n,q)$ and $\cL_{\rm st}(n,q)$. 
%
By the above, 
$\cL(n,q)\cong \cS(n,q)$
and $\cL_{\rm st}(n,q)\cong \cS_{\rm st}(n,q)=\{x\rightarrow q^ix+a \pmod n\mid a\in \bbZ_n, i\in [m]\}$;
in particular, 
we note that 
$\cL_{\rm st}(n,q)$ has size $nm$.
For missing proofs and more details, we refer to~\cite{cycl-subm1,cycl-subm2}.

\section{\label{LSlrs}Linear recurring seqence subgroups}
%
%
In this section, we establish the relation between non-standard pairs and non-standard linear recurring relation subgroups \cite{bn2}. For more background on linear recurring sequence, see, e.g., \cite{cycl-auto} or~\cite{ln}.

A sequence $\bfs=s_0, s_1, \ldots$ in~$\clF_q$ is called an $m$th order 
{\em linear recurring sequence\/}  if it satisfies a (homogeneous) linear recurrence relation of the form
\beql{LErec} s_k = \gs_{m-1}s_{k-1}+\cdots + \gs_1 s_{k-m+1}+\gs_0 s_{k-m}
\eeql
for all integers~$k\geq m$,
where $m\geq1$, $\gs_0\in\bbF_q^*$, and 
$\gs_1, \ldots,\gs_{m-1}\in\bbF_q$.
The monic polynomial 
\beql{LEf} f(x) = x^m -\gs_{m-1}x^{m-1}-\cdots -\gs_1x-\gs_0\eeql
in $\bbF[x]$ with $f(0)=-\gs_0\neq 0$ is called the {\em characteristic polynomial\/} of the recurrence
relation~(\ref{LErec}) and a sequence~$\bfs$ that satisfies~(\ref{LErec}) is called an {\em $f$-sequence\/}. 
Such a sequence is necessarily periodic, and we denote the (smallest) period of an~$f$-sequence~$\bfs$ by~$\per(\bfs)$. We say that an $f$-sequence~$\bfs$ in~$\clF_q$ is {\em cyclic\/} if there exists $\ga\in \clF_q$ such that $s_{k+1}/s_k=\ga$ for all $k\geq 0$; note that then, necessarily, $f(\ga)=0$.
Recall that $\xi\in \bbF_{q^m}$ denotes an element of order~$n$, and $\cU_{n,q}=\Gxi$ is the group of $n$-th roots of unity. 
If there exists an $f$-sequence $\bfs$ in~$\clF_q$ with $\per(\bfs)=n$ such that $\cU_{n,q}=\{s_0, s_1, \ldots, s_{n-1}\}$, then we say that $\cU_{n,q}$ is 
an~{\em $f$-subgroup\/}, and that $\cU_{n,q}$ is {\em represented\/} by~$\bfs$. Sometimes, such a group is referred to as a {\em linear recurring sequence subgroup\/} \cite{bn1}. 
A rather uninteresting way for $\cU_{n,q}$ to be an~$f$-subgroup is when $f$ is the minimal polynomial of~$\xi$ over~$\bbF_q$, representing $\cU_{n,q}$ by the cyclic $f$-sequence~$\bfs$ with $s_k=\xi^k$ ($k\geq0$). But sometimes there exist {\em non-cyclic\/} $f$-sequences that represent~$\cU_{n,q}$; in that case, if $f$ is a polynomial over~$\bbF_q$, we refer to $\cU_{n,q}$ as a {\em non-standard $f$-subgroup\/} or a {\em non-standard linear recurring sequence subgroup (NSLRS-group)\/} over~$\bbF_q$.
 
In this paper, we will be mainly interested in the case where the characteristic polynomial of the linear recurrence relation is {\em irreducible\/} over~$\bbF_q$. 
Our approach is based on the following results. For precise proofs, we refer to~\cite{cycl-subm1}.
\btm{LTgseq}\rm Let $g$ be irreducible over~$\bbF_q$ of degree~$m$, and let $\xi$ be a zero of~$g$ of order~$n$ in~$\bbF_{q^m}$, so that $m=\ord_n(q)$. Then a sequence $\bfs$ is a $g$-sequence in~$\bbF_{q^m}$ if and only if there are $L_0,\ldots, L_{m-1}\in\bbF_{q^m}$ such that
\beql{LEseq}s_k=L_0 \xi^k+L_1\xi^{qk}+\cdots + L_{m-1}\xi^{q^{m-1}k} \qquad (k\geq 0).\eeql
\etm
\bpf (Sketch)
Since $\xi, \xi^q, \ldots, \xi^{q^{m-1}}$ are all zeros of~$g$, sequences of the form as in the theorem are indeed $g$-sequences. A dimension argument shows that these are indeed all the $g$-sequences over~$\bbF_{q^m}$. 
\epf
\begin{teor}\label{Tirr} With the assumptions as in~\Tm{LTgseq}, 
a $g$-subgroup $\cU$ in an extension of~$\bbF_q$ is {\em unique\/}, and has the form $\cU=\Gxi=\cU_{n,q}$.
\end{teor}
\bpf
(Sketch) All zeros $\xi^{q^i}$ ($i\in[m]$) of~$g$ have the same period~$n$, and so every non-zero $g$-sequence $\bfs$ has minimal period $\per(\bfs)=n$.
\epf
For a generalization of this result to $f$-subgroups for general~$f$, and for additional background and references, see~\cite{cycl-auto}.
%
%
If we now combine the above results, we obtain the following.
\btm{LTFql}\rm With the assumptions as in~\Tm{LTgseq}, 
a group $\cU$ is a $g$-subgroup if and only if $\cU=\cU_{n,q}=\Gxi$.
In addition, there is a one-to-one correspondence between $g$-sequences representing $\cU_{n,q}$ and 
$\bbF_q$-linear maps $L\in\cL(n,q)$,
where $\bfs$ corresponds to~$L$ if~$s_k=L(\xi^k)$ ($k\geq0)$; moreover, $\bfs$ is cyclic if and only if the corresponding map $L$ is in~$\cL_{\rm st}(n,q)$.  
\etm
\bpf (Sketch) It is well-known that there is a one-to-one correspondence between $\bbF_q$-linear maps on~$\bbF_{q^m}$ and {\em $q$-polynomials\/} on~$\bbF_{q^m}$, maps of the form $L(x)=L_0x+L_1x^q+\cdots +L_{m-1}x^{q^{m-1}}$ with $L_0, L_1,\ldots, L_{m-1}\in \bbF_{q^m}$.  Now the result essentially is a direct consequence of the expression~(\ref{LEseq}) for $g$-sequences in~$\bbF_{q^m}$.
\epf
As a consequence, the notion of a non-standard pair $(n,q)$ coincides with that of a non-standard $g$-subgroup over~$\bbF_q$ with~$g$ irreducible of order~$n$.  
\section{\label{LSauto}Automorphisms of cyclic codes}
A linear $[n,k]_q$-code $C$ is an $\bbF_q$-linear $k$-dimensional subspace of~$\bbF_q^n$. We sometime refer to vectors in~$C$ as {\em code words\/}. The dual code $C^\perp$ of~$C$ is the collection of all vectors $\bfaa\in \bbF_q^n$ for which 
$(\bfaa, \bfc):=a_0c_0+\cdots +a_{n-1}c_{n-1}=0$ for all~$\bfc\in C$. A permutation $\pi\in \cS_n$ induces a permutation on~$\bbF_q^n$ by mapping a vector $\bfc=(c_0, c_1, \ldots, c_{n-1})$ to the vector 
\[\bfc^\pi=(c_{\pi^{-1}(0)}, c_{\pi^{-1}(1)}, \ldots, c_{\pi^{-1}(n-1)}).\] 
The group of permutation automorphisms of~$C$, denoted by~$\PAut(C)$, is the collection of all permutations $\pi\in \cS_n$ with the property that if $\bfc\in C$, then $\bfc^\pi\in C$. For later use, we remark that $\PAut(C^\perp)=\PAut(C)$ \cite[Lemma 1.3, (i)]{hufhand}.
The {\em cyclic shift\/} is the permutation $\gs=(0,1,\ldots, n-1) \in\cS_n$, mapping $i$ to $i+1\pmod n$ ($i\in[n])$.
A linear code $C\subseteq \bbF_q^n$ is {\em cyclic\/} if $\gs\in \PAut(C)$. By identifying a vector $\bfc=(c_0, c_1, \ldots, c_{n-1})\in \bbF_q^n$ with the polynomial $c(x)=c_0+c_1x+\cdots +c_{n-1}x^{n-1}$ in~$\cR_{n,q}:=\bbF_q[x] \bmod x^n-1$, a code of length~$n$ over~$\bbF_q$ corresponds to a subset of~$\cR_{n,q}$. A linear code is cyclic if and only if the corresponding subset is an {\em ideal\/} in~$\cR_{n,q}$. Now any ideal of~$\cR_{n,q}$ is {\em principal\/}, that is, generated by a unique monic polynomial. If $C=(g(x))$ is a cyclic code of length~$n$, then $g(x)$ is called the {\em generator polynomial\/} of~$C$ and $h(x)=(x^n-1)/g(x)$ is called the {\em parity-check polynomial\/} of~$C$. 

Let $\xi\in\bbF_{q^m}$ have order~$n$ and degree~$m=\ord_n(q)$ over~$\bbF_q$.
Let $\Tr(x)=x+x^q+\cdots +x^{q^{m-1}}$ denote the trace of~$\bbF_{q^m}$over~$\bbF_q$.
The code $C_{n,q}$ consisting of all code words
\[ \mu(a)=(\Tr(a\xi^0), \Tr(a\xi^1), \ldots, \Tr(a\xi^{n-1}))\]
with $a\in \bbF_{q^m}$ is called the {\em irreducible\/} (or {\em minimal\/}) cyclic code of length~$n$ over~$\bbF_q$. Note that $C_{n,q}$ is unique up to a permutation, and has dimension~$m$. 
The dual $C_{n,q}^\perp$ of~$C_{n,q}$ has as generator polynomial the minimal polynomial $g(x)$ of~$\xi$ over~$\bbF_q$, which has degree~$m$; such a code is called {\em maximal\/} \cite{mcws}. 
It is well-known and easily verified that beside the cyclic shift $\gs$, also the Frobenius mapping $\phi: x\rightarrow qx \pmod n$ (considered as an element of~$\cS_n$) is an automorphisms of~$C_{n,q}$ (to see this, note that if 
$\bfc\in\bbF_q^n$, then $c(x)^q=c(x^q)=c^{\phi}(x)$ corresponds to~$\bfc^\phi$). We write $\PAut_{\rm st}(C_{n,q})$ to denote the  group $\la \gs, \phi\ra$ generated by~$\gs$ and~$\phi$; note that $|\PAut_{\rm st}(C_{n,q})|=nm$. Recall that, by definition, $C_{n,q}$ is a NSIC-code if and only if $\PAut(C_{n,q})\supsetneq\PAut_{\rm st}(C_{n,q})$. 
The following result shows some unexpected connections between the concept of a NSIC-code
and various other notions of
non-standardness.
\btm{LTirrc}\rm An irreducible cyclic code $C_{n,q}$ is a NSIC-code if and only if $(n,q)$ is a non-standard pair.
More precise, we have that $\PAut(C_{n,q})=\PAut(C_{n,q}^\perp)=\cS(n,q)$ and $\PAut_{\rm st}(C_{n,q})=\PAut_{\rm st}(C_{n,q}^\perp)=\cS_{\rm st}(n,q)$. 
\etm 
\bpf
First, let $L\in \cL(n,q)$, with $\Psi(L)=\pi\in \cS_n$. Recall that  $C_{n,q}^\perp$ is the cyclic code with defining zero~$\xi$. So if $\bfc\in C_{n,q}^\perp$, then $c(\xi)=0$, hence 
\[0=L(0)=L(\sum_{i=0}^{n-1}c_i\xi^i)=\sum_{i=0}^{n-1}c_i\xi^{\pi(i)}=c^\pi(\xi),\]
and we conclude that $\bfc^{\pi} \in C_{n,q}^\perp$. Since $\bfc\in C_{n,q}^\perp$ was arbitrary, $\pi\in \PAut(C_{n,q}^\perp)$.

On the other hand, let $\pi\in \PAut(C_{n,q}^\perp)$. Since $(1,\xi, \ldots, \xi^{m-1})$ is a basis for~$\bbF_{q^m}$ over~$\bbF_q$, we can define an $\bbF_q$-linear map~$L$ on~$\bbF_{q^m}$ by setting $L(\xi^i)=\xi^{\pi(i)}$ for $i=0, \ldots, m-1$, and then extending $L$ by~$\bbF_q$-linearity. We claim that $L\in \cL(n,q)$ with $\Psi(L)=\pi$. To see this, let $j\in\{m, \ldots, n-1\}$. There are  $a_0, \ldots, a_{m-1}\in \bbF_q$ such that $\xi^j=a_0+a_1\xi+\cdots +a_{m-1}\xi^{m-1}$. This has two consequences. First, by definition of~$L$, 
\beql{LELexp}L(\xi^j)=a_0\xi^{\pi(0)}+\cdots +a_{m-1}\xi^{\pi(m-1)}.\eeql
Next, the vector
$\bfc=(a_0, \ldots, a_{m-1}, 0, \ldots, 0, -1, 0, \ldots, 0)$, with the entry -1 in position~$j$, has $c(\xi)=0$, so $\bfc\in C_{n,q}^\perp$, and since $\pi\in\PAut(C_{n,q}^\perp)$, we have $\bfc^\pi\in \PAut(C_{n,q}^\perp)$, that is,
\beql{LEcexp} 0=\sum_{i=0}^{n-1}c_i\xi^{\pi(i)}=a_0\xi^{\pi(0)}+\cdots +a_{m-1}\xi^{m-1} -\xi^{\pi(j)}.\eeql
Combining (\ref{LELexp}) and~(\ref{LEcexp}), we conclude that $L(\xi)=\xi^{\pi(j)}$. Since $j$ was arbitrary, we conclude that $L(\xi^j)=\xi^{\pi(j)}$ for all $j\in [n]$,
that is, $\Psi(L)=\pi$.
\epf
\section{\label{LSlex}Lifting and extension}
In this section, we discuss two methods to create new non-standard pairs from existing ones.

Let $\xi$ have order~$n$ and degree~$m$ over~$\bbF_q$, and let~$t$ be a positive integer for which~$\gcd(m,t)=1$. 
Since $\gcd(n,q^{ti}-1)=\gcd(n,q^{ti}-1, q^m-1)=\gcd(n,q^{\gcd(m,ti)}-1)=\gcd(n,q^{\gcd(m,i)}-1)$, we have $m=\ord_n(q)=\ord_n(q^t)$, hence the minimal polynomial $g(x)$ of~$\xi$ over~$\bbF_q$ is also the minimal polynomial of~$\xi$ over~$\bbF_{q^t}$. So in view of the results in~\Sec{LSlrs}, the following result is not too surprising.
\btm{LTlift}\rm ({\em Lifting\/}) Let $m$ and~$t$ be as above.
Then 
$\cS(n,q)=\cS(n,q^t)$ and $\cS_{\rm st}(n,q)=\cS_{\rm st}(n,q^t)$. So $(n,q)$ is non-standard if and only if $(n,q^t)$ is non-standard.
\etm
\bpf (Sketch)
Let $L\in \cL(n,q)$, so let $L(x)=\sum_{i=0}^{m-1}L_ix^{q^i}$ on~$\bbF_{q^m}$ with $L_i\in\bbF_{q^m}$ ($i\in [m]$), where we consider the indices modulo~$m$. Define $\tL(x)=\sum_{i=0}^{m-1} L_{it}x^{q^{ti}}$  ($x\in \bbF_{q^{mt}}$). Then $L=\tL$ on~$\bbF_{q^m}$, and $\tL$ is $\bbF_{q^t}$-linear on~$\bbF_{q^{tm}}$. Moreover, since $\la \xi \ra\subseteq \bbF_{q^m}$, we have $\tL(\xi^i)=L(\xi^i)=\xi^{\pi(i)}$ for all~$i\in [n]$, where $\pi=\Psi(L)$, so $\tL\in\cL(n,q^t)$ and $\Psi(\tL)=\Psi(L)$. 
Conversely, if $\tL\in \cL(n,q^t)$, then $\tL$ is $\bbF_{q^t}$-linear, hence $\bbF_q$-linear, on~$\bbF_{q^{mt}}$, and since $\tL$ fixes $\la \xi\ra$ set-wise, $\tL$ also fixes $\bbF_q(\xi)=\bbF_{q^m}$ set-wise. So the restriction~$L$ of~$\tL$ to~$\bbF_{q^m}$ is in~$\cL(n,q)$.
The claims are now obvious.
\epf
We will refer to the operation of passing from a non-standard pair~$(n,q)$ to a non-standard pair~$(n,q^t)$ with $\gcd(m,t)=1$ as {\em lifting\/}. In coding terms, note that if $\tC$ is the $\bbF_{q^t}$-span of the code~$C_{n,q}$, then $\PAut(\tC)=\PAut(C_{n,q})$.  And we have $C_{n,q^t}\subseteq \tC$, with equality if and only if the parity-check polynomial~$g$ of~$C_{n,q}$ (and of~$\tC$) is also irreducible over~$\bbF_{q^t}$, which holds if and only if $\gcd(m,t)=1$. 

The above operation on a non-standard pair~$(n,q)$ changed the value of~$q$. Our next operation changes~$n$. 
Recall that if $\xi\in\bbF_{q^m}^*$ has order~$n$, then $d=\gd_q(n)=n/\gcd(n,q-1)$ is the smallest positive integer such that $\xi^d\in \bbF_q$; in addition, $n=de$ with $e=\gcd(n,q-1)$ and $\gcd(d,(q-1)/e)=1$. 
First, we prove a lemma.
\ble{LLGphi}\rm
Let $\xi\in \bbF_{q^m}$ have order~$n$ and $q$-order $\gd_q(\xi)=\gd_q(n)=d=n/\gcd(n,q-1)$, and let $\cG\leq \bbF_q^*$ 
with $\xi^d\in \cG$. Then $\cG\la \xi \ra \leq \bbF_{q^m}^*$ and $|\cG\la \xi \ra|=d|\cG|$. 
\ele
\bpf
Obviously, $\cG\Gxi\leq \bbF_{q^m}^*$.  Since~$\xi^d\in\cG$, every element of~$\cG\la \xi\ra$ has the form $g\xi^i$ with $g\in\cG$ and $i\in[d]$. By definition of the $q$-order, all these elements are distinct, hence $\cG\la \xi\ra$  has order~$d|\cG|$.
\epf
Now we have the following `extension'' result.
\btm{LText}\rm ({\em Extension\/})
Let $n=de$ with $e=\gcd(n,q-1)$ and $d=\gd_q(n)$, so that $\gcd(d,(q-1)/e)=1$. For a positive integer~$f$, we have $d=\gd_q(n)=\gd_q(nf)$ if and only if $f\mid (q-1)/e$. Let $f\mid (q-1)/e$ and let $\gth$ have order~$nf$ in an extension of~$\bbF_q$. Then $\gth\in\bbF_{q^m}^*$, and $\la \gth\ra \cL(n,q)\leq \cL(nf,q)$; 
in particular, if $(n,q)$ is non-standard and $f\mid d(q-1)/n$, then $(nf,q)$ is also non-standard.
\etm
\bpf
With the assumptions in the theorem, we have that $\gd_q(nf)=def/\gcd(def, q-1)=df/\gcd(df,(q-1)/e)=df/\gcd(f,(q-1)/e)=d$ if and only if $f\mid (q-1)/e$. Next, assume that $f\mid (q-1)/e$. Let $\xi\in \bbF_{q^m}^*$ have order~$n$, and let $\gth$ have order~$nf$. 
Now $\cG=\la \gth^d \ra \leq \bbF_q^*$ and $|\cG|=ef$, hence $\la\gth\ra=\cG\la \xi\ra$ by~\Le{LLGphi}. Finally, if $g\in \cG$ and~$L\in \cL(n,q)$, then $gL(\la \gth\ra)=gL(\cG \Gxi) \subseteq g\cG \Gxi\subseteq \Gth$, hence $gL\in\cL(n,q)$. Obviously, $\cL_{\rm st}(nf,q)=\la \gth\ra \cL_{\rm st}(n,q)$, so the last conclusion is immediate.
\epf
%
Extension is related to the following code construction technique.
Let $C$ be a cyclic code of length~$n$ over~$\bbF_q$, with parity-check polynomial $g(x)\in\bbF_q[x]$, so that $g(x)\mid x^n-1$, and let $\nu\in\bbF_q^*$ with $\nu^{nf}=1$. Let $\tC\subseteq \bbF_q^{nf}$ be the collection of words 
\[\tbc=(c_0, \nu c_1, \ldots, \nu^{n-1}c_{n-1}, \nu^nc_0, \ldots, 
\nu^{nf-1}c_{n-1})\]
with $\bfc=(c_0, \ldots, c_{n-1})\in C$. If $c(x)=a(x)(x^n-1)g(x)$ for some $a(x)\in\bbF_q[x]$ with $\deg(a)<\deg(g)$,  then $\tc(x)=c(\nu x)((\nu x)^{nf}-1)/((\nu x)^n-1)=a(\nu x)(x^{nf}-1)/g(\nu x)$, hence $\tC$ is a cyclic code of length~$nf$ over~$\bF_q$, with parity-check polynomial $(x^{nf}-1)/g(\nu x)$. Note that if $g$ is irreducible with zero $\xi$, then $C=C_{n,q}$, and if the zero $\xi/\nu$ of $g(\nu x)$ has order~$nf$, then $\tC=C_{nf,q}$. 
%
Remark also that if $\tnu$ has order~$f$, if $C_0$ is the cyclic code consisting of all scalar multiples of the word $(1, \tnu, \tnu^2, \ldots, \tnu^{f-1})$, and if $C_1$ is the code with code words $(c_0, \nu c_1, \ldots, \nu^{n-1}c_{n-1})$ with $(c_0, c_1, \ldots, c_{n-1})\in C$, then $C_1$ is equivalent to~$C$ and $\tC$ is equivalent to the product code $C_0\times C_1$. 
For more information on this product code construction and its relation with extension, we refer to~\cite{cycl-subm1}. 

In \cite{cycl-subm1} we show that multiple lifts and extensions can always be obtained by a {\em single\/} lift, followed by a {\em single\/} extension.
%
\section{\label{LSes}Equally-spaced polynomials}
A polynomial $f(x)\in \bbF_q[x]$ is called {\em equally-spaced\/} if it is of the form $f(x)=g(x^k)$ for some integer~$k\geq2$. Then $\deg(f)=k\cdot \deg(g)$ and if $g(0)\neq 0$, then $\ord(f)=k\cdot \ord(g)$ and $\ord_q(f)=k\cdot \ord_q(g)$ (see~\cite{cycl-subm1}). Equally-spaced polynomials form a rich source of non-standard examples \cite{bn4}.
\btm{LTesirr} Let $k\geq2$ be a positive integer with $\gcd(k,q)=1$, let $g(x)\in\bbF_q[x]$ be monic with~$g(0)\neq 0$, of order~$n\geq1$ and degree~$m\geq1$, and suppose that $f(x)=g(x^k)$ is irreducible over~$\bbF_q$. Then the pair $(nk,q)$ 
is non-standard 
except when $f(x)=x^2+1$.
\etm
For a proof, we refer to \cite{bn4} or~\cite{cycl-subm1}. In \cite{cycl-subm1} we also show that if $C$ is cyclic of length~$n$ over~$\bbF_q$, with generator polynomial $g(x)$, then the product code $\bbF_q^k\times C$ is again cyclic, with generator polynomial $g(x^k)$.

Note that under the conditions of \Tm{LTesirr}, the polynomial $f(x)=g(x^k)$ is irreducible over~$\bbF_q$ if and only if $\ord_{nk}(q)=mk$. The next theorem states when this occurs.
\btm{LTfirr} Let $k$, $g(x)$, $n$, and $m$ be as in \Tm{LTesirr}, so with $\gcd(nk,q)=1$ and with $g$ irreducible, so with $m=\ord_n(q)$. 
Let $P(n,q)$ be the set of numbers that have only prime factors $r$ 
for which $r|n$ and $\gcd(r,(q^m-1)/n)=1$. Then the polynomial $f(x)=g(x^k)$ 
 is irreducible if and only if 
$k\in P(n,q)$, with $4\not\,\mid k$ if $2\in P(n,q)$ and $n\equiv 2\bmod 4$. 
\etm
The attraction of~\Tm{LTfirr} is that it is {\em constructive\/}. For a proof, we refer to~\cite{cycl-subm1}. 

%

\section{\label{LScases}Known non-standard pairs and their codes}
The following non-standard pairs $(n,q)$ and corresponding irreducible (NSIC) and maximal cyclic codes are known to us.\\
1) Pairs $(n,q)$ with $n\geq5$ prime and $m=\ord_n(q)=n-1$. The corresponding codes are the {\em repetition codes\/} and their duals, the {\em even-weight codes\/}, with corresponding polynomials $f(x)=(x^n-1)/(x-1)$. Here, $\cL(n,q)=\cS_n$.\\
2) Pairs $(n,q)$ with $n=q^m-1$, where $m>2$ or $m=2, q>2$. The corresponding codes are  the {\em $q$-ary simplex codes\/} and their duals, the primitive BCH codes with designed distance 2, with as polynomials the {\em primitive\/} polynomials over~$\bbF_q$. Here, $\cL(n,q)=\GL(m,q)$.\\
3) The pair $(n,q)=(23,2)$, where $m=\ord_n(q)=11$, with corresponding code the (dual of the) {\em binary Golay code\/}, with group $\cL(23,2)\cong M_{23}$.\\
4) The pair $(n,q)=(11,3)$, where $m=5$, with corresponding codes the (dual of the) {\em ternary Golay code\/}, with group $\cL(11,3)\cong \PSL(2,11)$.\\
5) Pairs $(n,q)=(kn_0, q)$ with $m=\ord_n(q)=km_0$, where $n_0>1$ and~$k\geq2$ are integers with $k>2$ if $n_0=2$ and $m_0=\ord_{n_0}(q)$. 
(See~\Tm{LTfirr} for the conditions under which this is the case.)
The corresponding codes are $\bbF_q^k \times C_{n_0,q}$ and their duals, with group $\cL(n,q)\cong S_k\wr \PAut(C_{n_0,q})$ (wreath product), and with corresponding polynomials the irreducible {\em equally-spaced\/} polynomials $f(x)=g(x^k)$ with $g(x)\in \bbF_q[x]$ and $f(x)\neq x^2+1$. \\
%
6) In addition, for every non-standard pair $(n,q)$ with $m=\ord_n(q)$, we have non-standard pairs $(nf,q^t)$ for every positive integer~$t$ such that~$\gcd(t,m)=1$ and every positive integer~$f$ such that~$f \mid (q^t-1)/\gcd(n,q^t-1)$ that can be obtained from $(n,q)$ by~{\em lifting and extension\/} as described in~\Sec{LSlex}. 

Possibly, the above examples exhaust all possibilities. In the next section, we show this to hold for the case where $m=2$.
\section{\label{LSclass}Classification for $m=2$}
For degree $m=2$, we have the following result.
\btm{LTm=2}The non-standard pairs $(n,q)$ with $m=\ord_n(q)=2$ and $q$-order~$d=n/\gcd(n,q-1)$ are the ones listed in \Sec{LScases} for which $m=2$ and are the following.\\
1) Pairs $(n,q)$ where $n=2e>4$ with $e\geq1$ integer for which $e\mid q-1$, with both $q$ and $(q-1)/e$ odd (the equally-spaced case); here $d=2$ and  $\cL(2e,q)\cong \bbZ_e\wr S_2$ (wreath product).\\
2) Pairs $(n,q)$ where $n=q^2-1$ and~$q\geq3$ (the primitive case); here $d=q+1\geq4$ and $\cL(n,q)=\GL(2,q)$.
And in addition, 
all pairs $(nf,q^t)$ obtained from a non-standard pair 
$(n=q^2-1,q)$ as above
by lifting and extension, so with $t$ odd and $f \mid (q^t-1)/(q-1)$; here $d=q+1$ and 
$\cL(f(q^2-1),q^t)=\la \gth\ra \GL(2,q)$, where $\gth\in \bbF_{q^t}^*$ has order~$f(q^2-1)$.  
\etm
\bpf (Sketch) If $d=2$, then the corresponding polynomial is of the form $x^2-\gs$ with $\gs\in\bbF_q^*$ and is irreducible over~$\bbF_q$.  A simple analysis, using \Tm{LTfirr}, 
quickly leads to the non-standard pairs as listed in case 1). 
Next, let $\xi\in\bbF_{q^2}$ be an element of degree~$m=2$, order~$n$, and $q$-order~$d>2$, so with minimal polynomial $g(x)=x^2-\gs_1 x-\gs_0$, where $\gs_0,\gs_1\in \bbF_q^*$. 
Let $T\in \cL(n,q)$ be the ``cyclic shift'' map defined by $T(x)=\xi x$ on~$\bbF_{q^2}$, 
and let $L\in \cL(n,q)$ with $L(1)=1$ and $L(\xi)=\gw +\nu \xi$, say. We proceed in several steps.\\
1. (Normalization) The element $\txi=\xi/\gs_1$ has minimal polynomial $\tg(x)=x^2-x-\gl$, where $\gl=\gs_0/\gs_1^2$. Obviously, the $q$-order of~$\txi$ again equals~$d$.
Writing $\tgw=\gw/\gs_1$, we have $T(1)=\gs_1 \txi, T(\txi)=\gs_1 \txi^2=\gs_1(\txi+\gl)$ and $L(1)=1, L(\txi)=\nu \txi+\tgw$. Put $D=\diag(1, \gs_1^{-1})$ and define 
\[\tL=L^D=
\left[\begin{array}{cc}
1&\tgw\\
0&\nu
\end{array}
\right], \qquad 
\tT=\gs_1^{-1}T^D=
\left[\begin{array}{cc}
0&\gl\\
1&1
\end{array}
\right].
\]
Note that $\tL$ and $\tT$ are the matrices of the maps $L$ and $\gs_1^{-1}T$ on $\bbF_{q^2}$, with respect to the basis $(1,\txi)$. \\
2. (Subgroup of $\PGL(2,q)$) Next, we identify the points of the 1-dimensional projective geometry $\PG(1,q)$ over~$\bbF_q$ with the sets $\bbF_q^* \ga$ ($\ga\in\bbF_{q^2}^*$), and we consider the group $\cG=\la \tT,\tL\ra$ generated by the matrices $\tT$ and~$\tL$ as a subgroup of~$\PGL(2,q)$.
Since $\txi^d\in\bbF_q^*$, every power $\txi^j$ of~$\txi$ is of the form $a\txi^i$ for some $i\in\{0, 1, \ldots, d-1\}$ and some $a\in\bbF_q^*$, hence $\cG\leq \PGL(2,q)$ fixes the set $O=\{1, \xi, \ldots, \xi^{d-1}\}$, considered as a subset of~$\PG(1,q)$, set-wise.\\
3. (Subgroup structure of~$\PGL(2,q)$) Now we use the known subgroup structure of~$\PGL(2,q)$, see, e.g., \cite{dick}, \cite{fa12}, and a separate analysis for the cases $3\leq d\leq 5$, to conclude that either~$L$ is standard, or $\cG$ is conjugate in~$\PGL(2,q)$ to one of its subgroups $\PGL(2,q_0)$ or~$\PSL(2,q_0)$.  Assuming the last case, we can then show that $d=q_0+1$, where $q=q_0^t$ with $t$ odd, and $\gl, \tgw, \nu\in \bbF_{q_0}$
with $\bbF_{q_0}=\bbF_p(\gl)$,
hence $\txi\in \bbF_{q_0}^2$.\\
4. As a consequence, $L$ fixes $\bbF_{q_0^2}$, and a simple analysis reveals that the restriction of~$L$ to~$\bbF_{q_0^2}$ is non-standard if $L$ itself is non-standard. Moreover, since $L$ fixes both $\bbF_{q_0^2}^*$ and $\la \xi\ra$ set-wise, it also fixes the intersection $\bbF_{q_0^2}^*\cap \la \xi\ra =\la \xi^\gd\ra$, where $\gd$ is the $q_0$-order of~$\xi$. So, writing $\gth=\xi^\gd$, we conclude that the group $\la \gth \ra$ is non-standard over~$\bbF_{q_0}$, as witnessed by the non-standard map~$L$ on~$\bbF_{q_0^2}$. \\
5. We have shown that $\gth$ is non-standard over~$\bbF_{q_0}$, with (maximal) $q_0$-order $q_0+1$. Now by \cite[Theorem~2.4]{bn3}, $\gth$ is primitive in~$\bbF_{q_0^2}$. \\
6. Finally, it can be shown that, as expected, the non-standard group~ $\la \xi\ra$ over~$\bbF_q$ can be obtained for the non-standard group~$\la \gth\ra$ over~$\bbF_{q_0}$ by lifting and extension.
\epf
In \cite{cycl-subm2} we shown that a nonstandard element $\xi$ of order~$n$ and degree~$m=\ord_n(q)$ over~$\bbF_q$ with maximal $q$-order $d=(q^m-1)/(q-1)$ is necessarily primitive, so has order $n=q^m-1$, thus generalizing \cite[Theorem~2.4]{bn3} for $m=2$ to {\em all\/}~$m$. The proof of the generalization is completely different and uses the recently completed classification of 1/2-transitive linear groups~\cite{lps}. We take this opportunity to remark that this result {\em does not\/} follow from \cite[Theorem 2.3]{zan} since the given proof is only valid when the subgroup $G$ in that theorem is {\em proper\/}.
\section{\label{LSdis}Discussion and conclusions}
Let $n$ and~$q$ be positive integers with~$\gcd(n,q)=1$, and let $m$ be the multiplicative order of $q$ molulo~$n$, the smallest positive integer such that $n \mid q^m-1$.
We say that the pair $(n,q)$ is {\em non-standard\/} if the collection~$\cL(n,q)$ of $\bbF_q$-linear maps on~$\bbF_{q^m}$ that fixes the group $\cU_{n,q}$ of $n$-th roots of unity set-wise contains maps not of the form $x\rightarrow cx^{q^j}$ with $c\in \bbF_{q^m}$ and $j\in[m]$.
In this paper, we have first linked this notion to that of {\em non-standard linear recurring sequence subgroups\/} \cite{bn1,bn2,bn3, bn4}. Then we showed that a pair $(n,q)$ is non-standard precisely when an irreducible cyclic code 
of length~$n$ over~$\bbF_q$ 
has ``extra'' permutation automorphisms (others than those generated by the cyclic shift and some Frobenius mapping); in this paper, we refer to such codes as {\em NSIC-codes}. 
%
A result by Brison and Nogueira from~\cite{bn3} for the case $m=2$ states that a non-standard pair $(n,q)$ with (maximal) $q$-order $d=\gd_q(n)=q+1$ is necessarily primitive, so has $n=q^2-1$.
Using the known subgroup structure of~$\PG(2,q)$  in combination with this result, we have
finished the classification of the non-standard pairs and NSIC-codes 
for the case of dimension~$m=2$
that was initiated by Brison and Nogueira in~\cite{bn1, bn2, bn3}.

Substantial information is available on the subgroup structure of~$\PGL(m,q)$ for $m=3$, and to a lesser extend also for~$m=4$, see, e.g., \cite{king}. 
By using similar methods, in combination with 
our generalization of~\cite[Theorem~2.4]{bn3} 
to {\em all\/}~$m$ in~\cite{cycl-subm2},
we expect that classification is also possible for the case $m=3$, and perhaps even for the case~$m=4$.

%
\section*{Acknowledgment}
I want to thank Karan  Khathuria, Vitaly Skachek, and Ludo Tolhuizen for proofreading and discussions. 
%

\end{document}